%% file: 00_main.tex
\documentclass{vgtc}                          % final (conference style)
\ifpdf%                                % if we use pdflatex
  \pdfoutput=1\relax                   % create PDFs from pdfLaTeX
  \usepackage{graphicx}                % allow us to embed graphics files
  \DeclareGraphicsExtensions{.pdf,.png,.jpg,.jpeg} % for pdflatex we expect .pdf, .png, or .jpg files
\else%                                 % else we use pure latex
  \usepackage{graphicx}                % allow us to embed graphics files
  \DeclareGraphicsExtensions{.eps}     % for pure latex we expect eps files
\fi%

\usepackage{microtype}                 % use micro-typography (slightly more compact, better to read)
\PassOptionsToPackage{warn}{textcomp}  % to address font issues with \textrightarrow
\usepackage{textcomp}                  % use better special symbols
\usepackage{mathptmx}                  % use matching math font
\usepackage{times}                     % we use Times as the main font
\usepackage{cite}                      % needed to automatically sort the references
\usepackage{tabu}                      % only used for the table example
\usepackage{booktabs}                  % only used for the table example
\usepackage{hyperref}
\usepackage{color}
\usepackage{enumitem}
\usepackage{graphicx}
\usepackage{fixltx2e}

\usepackage{multicol}
\usepackage{amsmath}
\usepackage{amssymb}
\usepackage{amsfonts}
\usepackage[font={scriptsize,sf}]{subfig}
\usepackage{floatrow}
\usepackage{bm} % for bold for vector
\usepackage{helvet}
\usepackage{algorithm}
\usepackage[noend]{algcompatible}
\usepackage{soul}
\usepackage{tikz}
\usepackage{pgfplots}
\usepackage{siunitx} % for SI unit
\usepackage{paralist} % for compact item

\usepackage[font={scriptsize,sf}]{caption}
\usepackage{float}
\floatstyle{plaintop}
\restylefloat{table}
\usepackage{algorithm}
\usepackage[noend]{algcompatible}
\usepackage{mdwlist}
 
\onlineid{9280}

\vgtccategory{Research}

\vgtcinsertpkg

\title{A Visual Analytics Framework for Reviewing Streaming Performance Data
\vspace{-5pt}}

\author{
    \hspace*{3pt}Suraj P. Kesavan\thanks{e-mail: \{spkesavan, kelli, tfujiwara, klma\}@ucdavis.edu}\\
    \scriptsize \hspace*{3pt}University of California, Davis
\and 
    \hspace*{-2pt}Takanori Fujiwara\footnotemark[1]\\
    \scriptsize University of California, Davis
\and 
    \hspace*{2pt}Jianping Kelvin Li\footnotemark[1]\\
    \scriptsize \hspace*{-2pt}University of California, Davis
\and 
    \hspace*{-5pt}Caitlin Ross\thanks{e-mail: rossc3@rpi.edu, chrisc@cs.rpi.edu}\\
    \scriptsize Rensselaer Polytechnic Institute
\and
    \hspace*{0pt}Misbah Mubarak\thanks{e-mail: mmubarak@anl.gov, rross@mcs.anl.gov}\\
    \scriptsize \hspace*{0pt}Argonne National Laboratory
    \vspace{-12pt}
\and 
    \hspace*{-8pt} Christopher D. Carothers\footnotemark[2]\\
    \scriptsize \hspace*{-4pt}Rensselaer Polytechnic Institute
    \vspace{-12pt}
\and 
    \hspace*{-4pt}Robert B. Ross\footnotemark[3]\\
    \scriptsize \hspace*{-7pt}Argonne National Laboratory
    \vspace{-12pt}
\and 
    \hspace*{-3pt}Kwan-Liu Ma\footnotemark[1]\\
    \scriptsize \hspace*{4pt}University of California, Davis\hspace*{1pt}
    \vspace{-12pt}
}

\abstract{%
\input{0_abstract.tex}
} % end of abstract

\CCScatlist{
  \CCScatTwelve{Human-centered computing}{Visu\-al\-iza\-tion}{Visualization application domains}{Visual analytics}
  \vspace{-2pt}
}

\begin{document}


\definecolor{tf}{rgb}{0,0.6,0}
\newcommand{\takanori}[1]{{\color{tf}\textbf{[}TF\@: #1\textbf{]}}}

\newcommand \add[1]{{\textcolor{black}{#1}}}
\newcommand \del[1]{\textcolor{blue}{}}

\maketitle
\input{1_introduction.tex}
\input{2_relatedwork.tex}
\input{3_design_requirements.tex}
\input{4_instrumentation_module.tex}
\input{5_analysis_module.tex}
\input{6_visualization_module.tex}
\input{7_case_studies.tex}
\input{8_discussion.tex}
\input{9_conclusions.tex}

\acknowledgments{This research has been supported by U.S. Department of Energy through grant DE-SC0014917.}

% \clearpage
%\bibliographystyle{abbrv}
\bibliographystyle{abbrv-doi}

\bibliography{00_main}
\end{document}

%% file: 0_abstract.tex
Understanding and tuning the performance of extreme-scale parallel computing systems demands a streaming approach due to the computational cost of applying offline algorithms to vast amounts of performance log data.
Analyzing large streaming data is challenging because the rate of receiving data and limited time to comprehend data make it difficult for the analysts to sufficiently examine the data without missing important changes or patterns.
To support streaming data analysis, we introduce a visual analytic framework comprising of three modules: data management, analysis, and interactive visualization.
The data management module collects various computing and communication performance metrics from the monitored system using streaming data processing techniques and feeds the data to the other two modules.
The analysis module automatically identifies important changes and patterns at the required latency.
In particular, we introduce a set of online and progressive analysis methods for not only controlling the computational costs but also helping analysts better follow the critical aspects of the analysis results.
Finally, the interactive visualization module provides the analysts with a coherent view of the changes and patterns in the continuously captured performance data.
Through a multi-faceted case study on performance analysis of parallel discrete-event simulation, we demonstrate the effectiveness of our framework for identifying bottlenecks and locating outliers.
\vspace{-2pt}

%% file: 1_introduction.tex
\vspace{-2pt}
\section{Introduction}
High-performance computing (HPC) systems, such as massively parallel supercomputers, are essential to solving complex computational problems in diverse fields from weather forecasting and earthquake modeling to molecular dynamics and drug design.
To ensure the efficiency of HPC applications and maximize the utilization of supercomputer systems, performance monitoring and optimization are essential tasks.
As performance logs obtained from such large-scale systems tend to be large and complex, adequate visual analytics tools are needed to interpret and reason the data.

With computational abilities reaching exascale computing (i.e., capable of performing $10^{18}$ FLOPS in the near future), adopting in-situ workflows~\cite{ma2007situ, Larsen:ISAV17} has become popular.
In-situ workflows integrate visualization into the simulation pipeline to steer the simulation by executing real-time performance analysis.
Therefore, performance visualization tools not only have to handle streams of real-time performance data but also implement visualizations that can help the analyst comprehend their underlying behaviors.
However, existing performance analysis techniques are constrained by several challenges.
First, from streaming data, a performance analyst needs to catch important patterns, changes, or anomalies in real-time (active monitoring~\cite{dasgupta2018human}) without missing them.
Furthermore, the analyst often wants to identify causal relations of an occurred phenomena  (situation awareness~\cite{dasgupta2018human}).
Because streaming performance data is continuously changing with high-volume and high-variety properties, without any algorithmic and visual supports, conducting the above analysis is almost infeasible.
While existing visual analytic systems~\cite{li2017visual, fujiwara2017visual} aim to support the analysis of dynamic performance data, support for real-time analysis is lacking.

To address the above challenges, we introduce a visual analytics framework for analyzing streaming performance data. 
We first identify the key analysis requirements through extensive exchange with HPC experts. 
The requirements include revealing temporal patterns from multivariate streaming data and correlate these temporal patterns to their network behaviors. 
To\,handle\,streaming\,data and fulfill the experts' requirements, our framework comprises (1) data management, (2) analysis, and (3) interactive visualization modules. 

To support the analysis and visualization of streaming performance data, we contribute a data management module for efficiently combining and processing data streams at interactive speed (e.g., 1 second \cite{turkay2017designing}).
Our data management module is also designed to be run as a subsystem in remote HPC or simulation systems to enable in-situ data processing and interactive visualizations.
It collects both multivariate time-series and network communication data, performs user-specified analytical processing, and delivers the results as data streams to the analysis and interactive visualization modules.

As for the analysis module, we design a set of algorithms for visually analyzing multivariate streaming data in real\add{-}time. 
In particular, we apply change point detection to identify key changes, time-series clustering and dimensionality reduction to reveal both common and outlier behaviors, and causal relation analysis to identify metrics' co-influences. 
A major challenge to develop these analysis methods is the constraint of the computation time. 
As new data keeps coming in, the computation must be fast enough to provide up-to-date results. 
To achieve this, we design our algorithms in either an online or a progressive manner. 
Online (or incremental) methods~\cite{albers2003online} calculate the new result by using the result from data already obtained as the base and then updating it according to the newly obtained data. 
Therefore, if applicable, online methods are suitable for streaming data analysis. 
On the other hand, progressive methods~\cite{turkay2018progressive} provide useful intermediate results for data analysis within a reasonable latency when the computational cost to complete an entire calculation is too high. 
While many progressive methods internally utilize online algorithms to generate the intermediate results, they may use a different approach, such as reducing computation with approximation~\cite{pezzotti2017approximated}.
Another challenge is that the results from the analysis may keep causing drastic changes in the visualization content, such as node positions obtained from dimensionality reduction.
Showing such results may disrupt the analyst's mental map. 
Therefore, we consider mental map preservation while designing our analysis methods.

The interactive visualization module is for the analyst to interpret the results from the analysis module.
The module provides a fully interactive interface to help relate the temporal behaviors from the historical context with the intermediate results to make critical observations. 
Additionally, we provide visual summaries for indicating when and what causes a particular performance bottleneck.
We demonstrate the efficiency and effectiveness of our framework with performance evaluation and a multi-faceted case study analyzing the data collected from a parallel-discrete event simulator.

%% file: 2_relatedwork.tex
\vspace{-1pt}
\section{Related Work}
We survey relevant works in performance visualization of parallel applications and streaming data visualization.

\subsection{Performance Visualization of Parallel Applications}
\label{sec:vaHPC}
%1. general VA
%2. network VA
%3. simulator (ROSS visualization)
%4. time-series ones (PacificVAST, etc)

Various visualizations have been developed to help performance analysis. 
Isaacs et al.~\cite{isaacs2014state} provided a comprehensive survey of existing performance visualizations. 
Here, we describe only visualizations for analyzing parallel applications.

One of the major research topics is visualizing the HPC system's network together with the performance metrics to analyze communication patterns.
Since the HPC system has many compute nodes (e.g., more than thousands of nodes) connected using a complex network topology (e.g., Dragonfly~\cite{dragonfly}), the researchers have introduced several advanced visualizations.
For example, Boxfish~\cite{landge2012visualizing} visualizes 3D torus networks by using a 3D mesh representation or by projecting the view onto a 2D plane and analyze network traffic along with their topological properties.
Similarly, Bhatele et al.~\cite{bhatele2016analyzing} analyzed Dragonfly-based networks using a radial layout and a matrix view to show inter-group and intra-group links between the compute nodes.
Fujiwara et al.~\cite{fujiwara2017visual} utilized node-link diagrams and the matrix-based representations with hierarchical aggregation techniques to generalize to any network topology. 
Li et al.~\cite{li2017visual} developed flexible visualization to analyze the network performance on the Dragonfly network by applying data aggregation techniques to scale for large scale networks.
While the works above visualized communication patterns, they do not provide sufficient methods for temporal analysis of performance behaviors. 

Several researchers have studied techniques to support temporal analysis. 
The Ravel visualization tool~\cite{isaacs2014combing} visualizes execution traces and event histories of parallel applications using logical time instead of physical time.
Using logical time allows the application developers to analyze the execution sequence from the program's perspective. 
Muelder et al.~\cite{muelder2016visual} introduced the ``behavior lines'' for analyzing cloud computing performance.  
These lines show an overview of the behavioral similarity of multivariate time-varying performance data.
To analyze the performance behaviors from the large-scale data, Fujiwara et al.~\cite{fujiwara2018visual, li2019visual} integrated advanced time-series analysis methods, including clustering, dimensionality reduction, and change point detection methods, into their visual analytics system.
In terms of coupling the advanced time-series analysis methods with visualizations, \cite{fujiwara2018visual} is the most closely related work. 
However, these existing methods including \cite{fujiwara2018visual} are not appropriate to analyze streaming performance data.
To analyze streaming performance data, we need to take additional considerations, such as calculation costs of analysis methods and cognitive loads of visualizations.

Only a few works have approached streaming performance data. 
For example, De Pauw et al.~\cite{de2013visualizing} developed a visualization system for real-time monitoring of a job allocation in shared computer resources.% (e.g., MapReduce~\cite{dean2008mapreduce} clusters). 
They used a streamgraph-like visualization which encodes time and a number of resources used by each job as $x$-direction and height of the stacked area, respectively.
Sanderson et al.~\cite{sanderson2018coupling} showed real-time performance data with basic visualizations, including line charts and heatmaps over the compute node layout.
However, these existing methods provide only limited information (e.g., currently used resources).
While these works only support analyzing one type of streaming performance data, our framework is designed for real-time analysis of both time-series and communication data streams.
Our framework also provides multiple coordinated views with interactive visualizations to analyzing both types of data streams together for real-time performance analysis and monitoring.

\subsection{Visualization for Streaming Data Analysis}
While visual analytics for streaming data has not been matured yet in performance visualization, researchers have developed visualizations in other domains, such as fraud detection~\cite{webga2015discovery} and factory management~\cite{xu2017vidx}.
Dasgupta et al.~\cite{dasgupta2018human} provided a comprehensive survey of streaming data visualization and its challenges. 
One of the major challenges with streaming data is how we show important changes or meaningful patterns with a low cognitive load since available data is constantly updated~\cite{dasgupta2018human,krstajic2013visualization}. 

A common approach taken is the simplification of visual results (e.g., aggregation).
For example, Xu et al.~\cite{xu2017vidx} developed extended Marey's graph for monitoring assembly line performance. 
While Marey's graph is originally used for train schedules, it suffers from visual cluttering when the data gets updated frequently. 
To solve this issue, the extended Marey's graph in \cite{xu2017vidx} emphasizes only the abnormal behaviors of the assembly line (e.g., causing delays of assembly processes) by aggregating the other behaviors.
Some researchers used dimensionality reduction (DR) methods to summarize changes in streaming multivariate data.
For instance, Cheng et al.~\cite{cheng2016framework} used the multidimensional-scaling (MDS) for showing an overview of similarities between temporal behaviors in streamed data. 
In addition to this, they introduced the concept of sliding MDS, which visualizes temporal changes in the similarities between selected points as line paths.
To detect the rating fraud in e-commerce stores, Webga and Lu~\cite{webga2015discovery} used singular-value decomposition to show the dissimilarities of the rating streams. 
However, when streaming data has high volume and velocity (i.e., frequent updates), the algorithms used in the works above could not work well due to the computational cost.
To handle high volume and velocity streaming data, the completion time of algorithms used for visual analytics should be constant and/or less than the data-collection rate even when a total number of data points or features are increasing~\cite{aggarwal2013survey}. 

Recently, progressive visual analytics~\cite{turkay2018progressive} is being actively studied.
Producing useful results with a latency restriction is a common requirement.
In fact, a few studies started to apply progressive visual analytics methods to handle streaming data.
For example, Pezzotti et al.~\cite{pezzotti2017approximated} developed Approximated t-SNE (A-tSNE) as a progressive version of t-SNE~\cite{maaten2008visualizing} which is a commonly-used non-linear DR method. 
They showed the usage of A-tSNE for streaming data visualization in their case study.
The work in \cite{fujiwara2019incremental} enhanced the existing incremental principal component analysis (PCA)~\cite{ross2008incremental} for streaming data visualization. 
In addition to the incremental calculation, their method maximizes the overlap of data points between previous and updated PCA results to keep the user's mental map.
We also use progressive methods to support the analysis of streaming \add{of} large performance data. 
However, while these two works above~\cite{pezzotti2017approximated,fujiwara2019incremental} focused only on handling new data points fed from streams, our methods can also be used when new features (values for new time points in our case) are obtained continuously.

%% file: 3_design_requirements.tex
\section{Background and Design Requirements}
\label{sec:bg_and_dr}
We describe the characteristics of streaming HPC performance data and design requirements identified through the collaboration with HPC experts. 
These lead to the design of our framework.

\subsection{Characteristics of Streaming Performance Data}
\label{sec:data}
Analysis of performance bottlenecks in HPC can be categorized into three key domains according to the HAC model~\cite{schulz2011interpreting} as follows: (a) Hardware, (b) Application, and (c) Communication domains. 
The hardware domain consists of network nodes and physical links between them; the application domain represents the physical or simulated system designed to solve the underlying problem; the communication domain represents a communication network, which captures the communication patterns of the application.
Since there exists a large number of visualizations that enable data analysis on the application domain using both post-hoc~\cite{biddiscombe2007time} and in-situ~\cite{ma2007situ,Larsen:ISAV17} based workflows, we especially focus on the data derived from the hardware and communication domains. 

Performance counters and other measurement devices on modern microprocessors record various metrics from the hardware domain at a uniform sampling rate.
The recorded data can be represented as a $d$-dimensional vector, where $d$ is the number of measured metrics.
Let $n$ be the number of entities (e.g., compute nodes, network routers, or MPI processes mapped to CPU cores) which provide $d$ measured metrics.
The resulting streaming data can be represented as the third-order tensor object of shape $n \times d \times t$, where $t$ is the number of samples recorded and is continuously growing with time.

Additionally, in general, $n$ entities communicate with each other to run \del{the }parallel applications. 
The communication data can be represented as a weighted graph where nodes correspond to the entities\add{,} and links represent the amounts of communications (e.g., message packet sizes) between the entities. 
Because communication bottlenecks often cause \del{the }performance bottlenecks, analyzing the performance metrics with the communication data can help the analyst in locating performance issues.

\subsection{Design Requirements}
\label{sec:requirements}
Our goal is to design a visual analytics framework that helps real-time analysis of streaming performance metrics and communication data.
Simultaneously analyzing both performance metrics and communications is critical because the communications heavily affect the performance\del{,} and vice versa. 
For example, a communication bottleneck in a network of compute nodes could be due to a large number of packets transferred between certain compute nodes leading to excessive wait times for the next computation.
To understand such specific analysis needs, we work in collaboration with HPC experts and enumerate the design requirements of the visual analytics framework. 
The design process followed a user-centered approach consisting of several discussions where we presented a prototype to the experts to probe further requirements\del{,} and modified the visualization framework, accordingly.
Below we describe the design requirements (\textbf{R1-–R5}) developed during these discussions.
These requirements comprehend analysis tasks of both active monitoring~\cite{dasgupta2018human} (\textbf{R1}, \textbf{R2}, and \textbf{R4}) and situational awareness~\cite{dasgupta2018human} (\textbf{R1--R5}).

\vspace{2pt}
\noindent\textbf{R1: Detect key changes that deviate from a baseline behavior.}
Because large-scale simulations are often long-running applications, it becomes impractical to \del{constantly }review performance metrics for changes \add{constantly}.
Hence, it is critical to automatically find when and where a change occurs in the behavior to narrow down the search space.

\vspace{2pt}
\noindent\textbf{R2: Enhance the interpretability to identify performance behavior patterns.}
As discussed in \autoref{sec:data}, the collected data contains time-series data continuously arriving from multiple processes.
From the streaming data, finding similar and dissimilar behaviors from many processes in real\add{-}time is extremely challenging because of its high-volume and high-velocity.
Therefore, we need to provide \del{a }functionality that helps the analyst identify the behavior patterns.

\vspace{2pt}
\noindent\textbf{R3: Derive causal relations among different performance metrics.}
A behavior change (e.g., excessive wait times) can trigger an undesired effect (e.g., low data receives). 
Therefore, the analyst would want to identify the causal relations among the multiple performance metrics.
Since it is difficult to monitor several metrics (e.g., five metrics) all at once, we aim to support the analysis of causal relations between metrics that significantly affect the performance.

\vspace{2pt}
\noindent\textbf{R4: Provide visualizations for analyzing communication patterns together with performance behaviors.}
As mentioned in \autoref{sec:data}, analyzing the performance data from both hardware and communication domains and relating these domains are important.
Thus, our framework should provide visualizations that help the analyst relate the performance behaviors to the communication patterns.% along the timeline.

\vspace{2pt}
\noindent\textbf{R5: Enable users to analyze the data interactively.}
Due to the complexity of HPC performance data, there are various aspects that the analyst wants to analyze while cooperating with his/her domain knowledge. 
The analytical process may proceed and change according to discovered patterns or findings. 
Therefore, it is essential to provide interaction methods that can be used during analysis.

%% file: 4_instrumentation_module.tex
\section{Visual Analytics Framework}
\label{sec:framework}
\begin{figure}[t]
    \captionsetup{farskip=0pt}% <--- no gap at the top
    \centering
    \includegraphics[width=\linewidth]{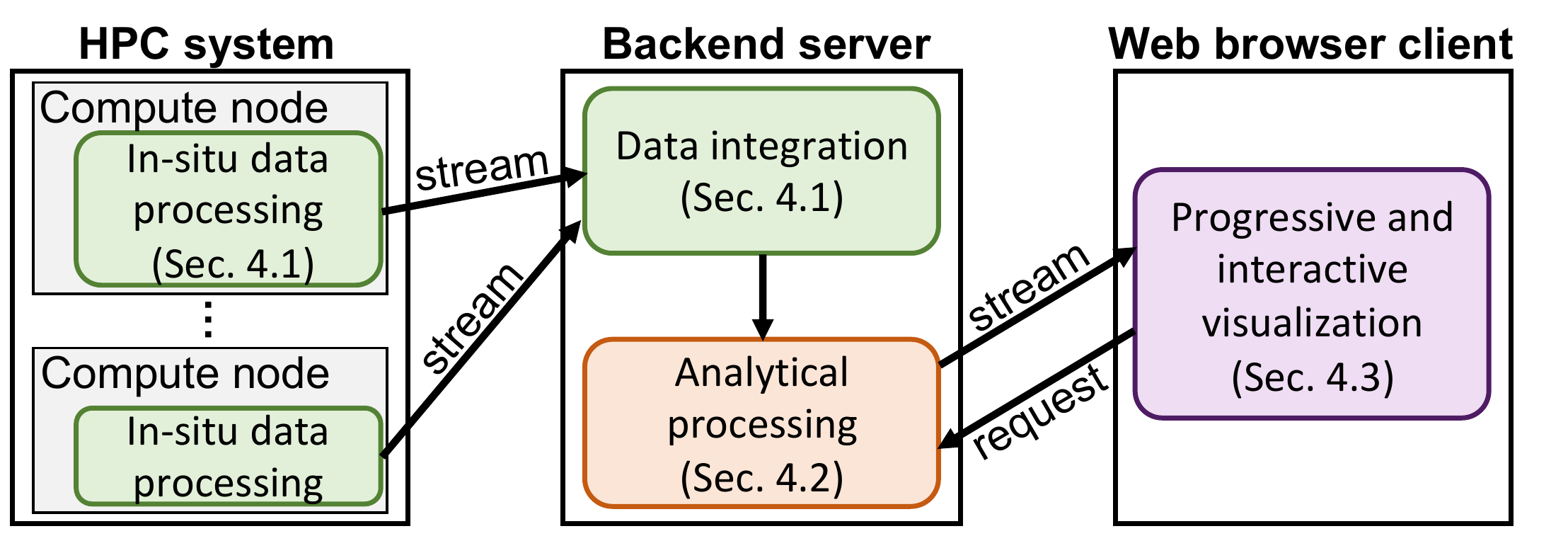}
    \caption{The overview of pipeline of our visual analytics framework. Data management module (green) joins and indexes the time-series and communication data collected from the HPC system, and then passes them to the analysis module (orange) for analytical processing. The results are then streamed to the interactive visualization module (purple) for rendering. Also, the analysis results can be updated based on interactions (e.g., changing algorithms' parameters).}
    \label{fig:framework}
\end{figure}

To fulfill the above requirements, we design our framework's pipeline.
The pipeline, as shown in~\autoref{fig:framework}, consists of (1) data management, (2) analysis, and (3) interactive visualization modules.
The entire framework is developed using C++ and Python for the back-end and JavaScript for the front-end.
The source code for a major portion of our framework is available in \cite{supp}.
Below we describe the methodology for each module in detail. 

\vspace{-5pt}
\subsection{Data Management Module}
For performance analysis using streaming data, our data management module is designed to leverage with \add{the} in-situ data processing and manage data flows in a server-client architecture, as shown in \autoref{fig:framework}.
Our framework follows the co-processing model for in-situ data processing described in~\cite{ma2007situ}. HPC applications or parallel simulations that adopt  this model can be seamlessly integrated into our framework for real-time analysis and visualization.
To collect performance data from HPC systems, the data management module uses agent programs that can be installed in multiple compute nodes.
Since the performance data on each compute node can be very large, directly collecting the raw data and streaming to the backend server of the visual analytics system requires high network bandwidth.
The minimum data granularity and resolution can be specified for in\add{-}situ processing of the time-series and communication data.
Common data transformations, such as sampling, filtering, and aggregation, can be specified in the in-situ data processing.
The results of in-situ data processing are streamed to the backend server, where the data management module performs data integration and indexing to allow the analysis module to perform analytical processing on the results effectively.
The analytical processing results are then streamed to the web browser for rendering visualizations.
The use of in-situ data processing greatly reduces network bandwidth, and thus allows our visual analytics framework to effectively facilitate real-time analysis and monitoring of streaming performance data.

%% file: 5_analysis_module.tex
\vspace{-3pt}
\subsection{Analysis Module}
\label{sec:algorithms}
To support the design requirements, the analysis module provides several automatic analysis methods. 
The difficulty in applying the algorithms to streaming data is that the computation time should be shorter than a time span of the data update. 
Thus, instead of using traditional offline algorithms, we introduce methods using an online or progressive approach.
Additionally, to make it easier to follow changes in the analysis results, we also provide algorithms to keep visual consistency between the previous and updated results. 
% Traditional non-streaming algorithms are effective for performing interactive analysis for stationary data, but do not work effectively for streaming data.
% The main reason is because the growth of data can become exponential making calculations become increasing expensive with time and eventually disrupting the user interactivity.
% To achieve this, we make use of \textit{online algorithms}, that are capable of processing data piece-by-piece, sequentially and ensure that computations are finished and associated visualizations are rendered within temporal limits.
% the simulation-induced latency and also perform visual interaction without delay. 

\vspace{-3pt}
\subsubsection{\hspace*{-3pt}Online\,Change\,Point\,Detection\,for\,Multiple\,Time-Series}
% \subsection{Progressive Change Point Detection}
\label{sec:cpd}
To address \textbf{R1}, we design an online change point detection (CPD) method.
As discussed in \autoref{sec:data}, for each metric, data obtained from HPC systems consist of $n$ multiple time-series~\cite{li2017visual,fujiwara2018visual,muelder2016visual} (e.g., message packets sent from each router). 
However, the existing online CPD methods are designed to identify change points for a single time-series. 
To apply CPD on multiple time-series data, we employ a similar approach to \cite{qahtan2015pca}.
Before using CPD, to obtain a single time-series, their method reduces multiple values to a ``representative'' value for each newly obtained time point with PCA.

\begin{figure}
    \captionsetup{farskip=0pt}% <--- no gap at the top
	\centering
	\hspace*{-8pt}
	\subfloat[Original streamed time-series]{
     \includegraphics[width=0.48\linewidth,height=0.21\linewidth]{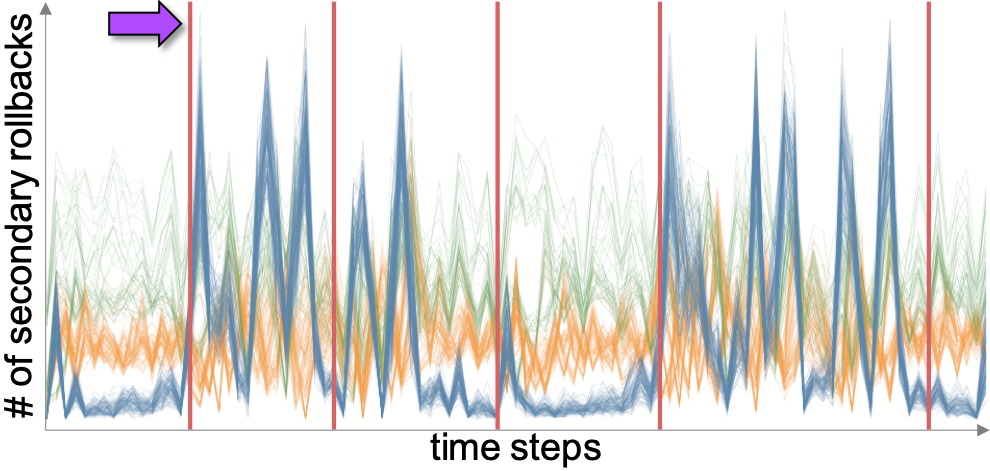}
     \label{fig:cpd_a}
    }
    \subfloat[With the ordinary PCA]{
     \includegraphics[width=0.48\linewidth,height=0.21\linewidth]{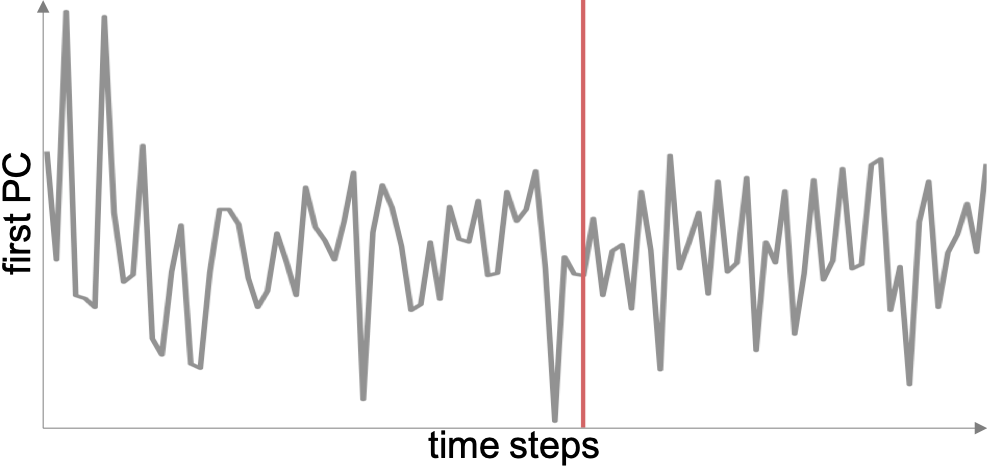}
     \label{fig:cpd_b}
    }
    \hspace*{-5pt}
    \\
    \vspace{1pt}
    \hspace*{-8pt}
    \subfloat[With incremental PCA (IPCA)]{
     \includegraphics[width=0.48\linewidth,height=0.21\linewidth]{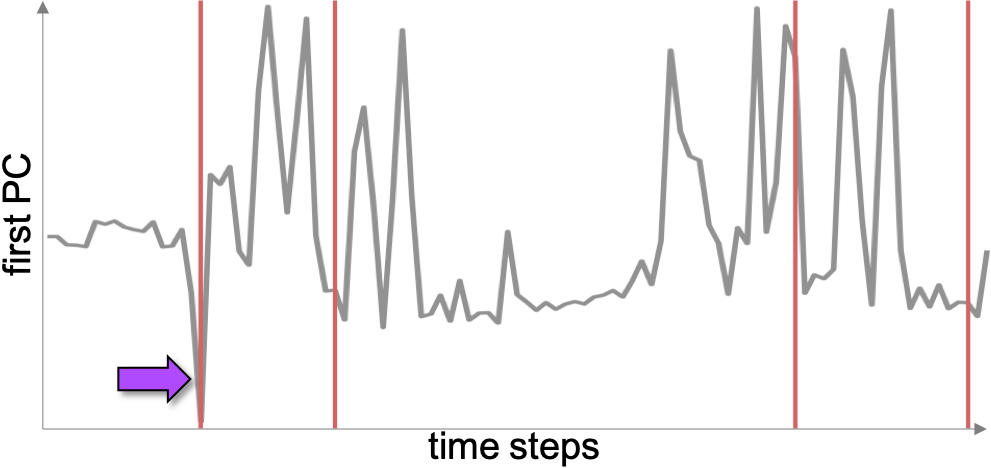}
     \label{fig:cpd_c}
    }
    \subfloat[With IPCA and sign adjustment]{
     \includegraphics[width=0.48\linewidth,height=0.21\linewidth]{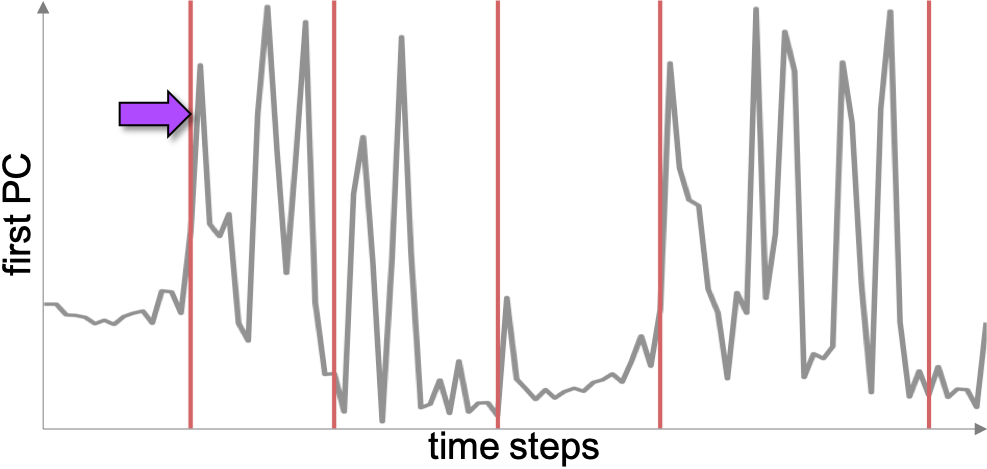}
     \label{fig:cpd_d}
    }
    \hspace*{-5pt}
   	\caption{
   	The comparison of change points detected with the different approaches. 
   	In (a), each $y$-coordinate represents the number of secondary rollback in each entity (refer to \autoref{sec:case_study_data} for details). 
   	The color of each polyline represents the cluster ID obtained with the method described in \autoref{sec:clutering}.
   	In (b), (c), and (d), the resultant single time-series with each approach are shown with the gray polylines.
   	The detected change points are also indicated with the red vertical lines (in (a), we show the same change points with (d)).}
	\label{fig:cpd}
\end{figure}

% inc PCA instead of PCA
While the approach in \cite{qahtan2015pca} applies the ordinary PCA~\cite{jolliffe1986principal}, we use incremental PCA (IPCA) of Ross et al.~\cite{ross2008incremental}.
By using IPCA, we can incrementally update the PCA's model when we obtain a new time point and then use this updated model for generating the representative value. 
This approach can capture the important information from multiple time-series with consideration of the variance in the past observations. 
\autoref{fig:cpd_a}, \ref{fig:cpd_b}, and \ref{fig:cpd_c} show the original streamed data comprising of 256 time-series, the result with the ordinary, and the result with IPCA, respectively. 
While \autoref{fig:cpd_b} does not summarize any useful patterns from the original data, we can see that \autoref{fig:cpd_c} tends to summarize the blue lines since these lines have higher variances over time compared with the others.

% sign flipping
However, (incremental) PCA causes an arbitrary sign flipping for each principal component (PC), as discussed in \cite{fujiwara2019incremental}. 
% \cite{bro2008resolving,jeong2009understanding,turkay2017designing,fujiwara2019incremental}. 
The arbitrary sign flipping may cause or hide the drastic changes in a PC value, and thus a CPD method may misjudge the flipping as the change point or overlook the change point.
To solve this issue, we generate coherent signs based on the cosine similarity of the PCs of the previous and current results. 
If the cosine similarity is smaller than zero, these PCs have opposite directions from each other, and thus, we flip the sign of the updated PCA's PC.  
\autoref{fig:cpd_c}, and \autoref{fig:cpd_d} show the results without and with the sign adjustment, respectively.
While the first peak in \autoref{fig:cpd_a}, as indicated with the purple arrow, is appeared as the negative peak in \autoref{fig:cpd_c}, \autoref{fig:cpd_d} shows the corresponding peak in the same direction as in \autoref{fig:cpd_a}. 

% AFF
The remaining process of the algorithm is applying an online CPD method that is designed to detect changes on a single streamed time-series.
We choose the method developed by Bodenaham and Adams~\cite{bodenham2017continuous} because, unlike most of the others~\cite{bifet2018machine}, their approach requires only one parameter.
Minimizing the number of required parameters is important, especially for streaming data, since parameter tuning is extremely difficult while the data is continuously changing and users are often unaware of the characteristics of the data (e.g., trend, patterns, and cycles) in advance.
% called the adaptive forgetting factor (AFF), which adaptively controls the effect from the past observations to the detection. 
The required parameter in \cite{bodenham2017continuous} is called the significance level $\alpha$ ($0 \leq \alpha \leq 1$), which controls a time-window width referred for CPD.
As $\alpha$ increases, the detector becomes more sensitive and also tends to generate more false detections~\cite{bodenham2017continuous}.
% \del{We set a default value of $\alpha=0.01$ which follows their default.}
% We set a default value of $\alpha=0.05$ while detecting for changes and the detected changes are shown as red vertical lines in \autoref{fig:cpd_a}.
We set a default value of $\alpha=0.01$. 
\autoref{fig:cpd_a} shows the result with our method. 
We can see that our method detects changes corresponding to the starts and ends of the high peaks.

\begin{figure}[tb]
    \captionsetup{farskip=0pt}% <--- no gap at the top
	\centering
	\subfloat[Without our cluster ID reassignment]{
     \includegraphics[width=0.95\linewidth,height=0.24\linewidth]{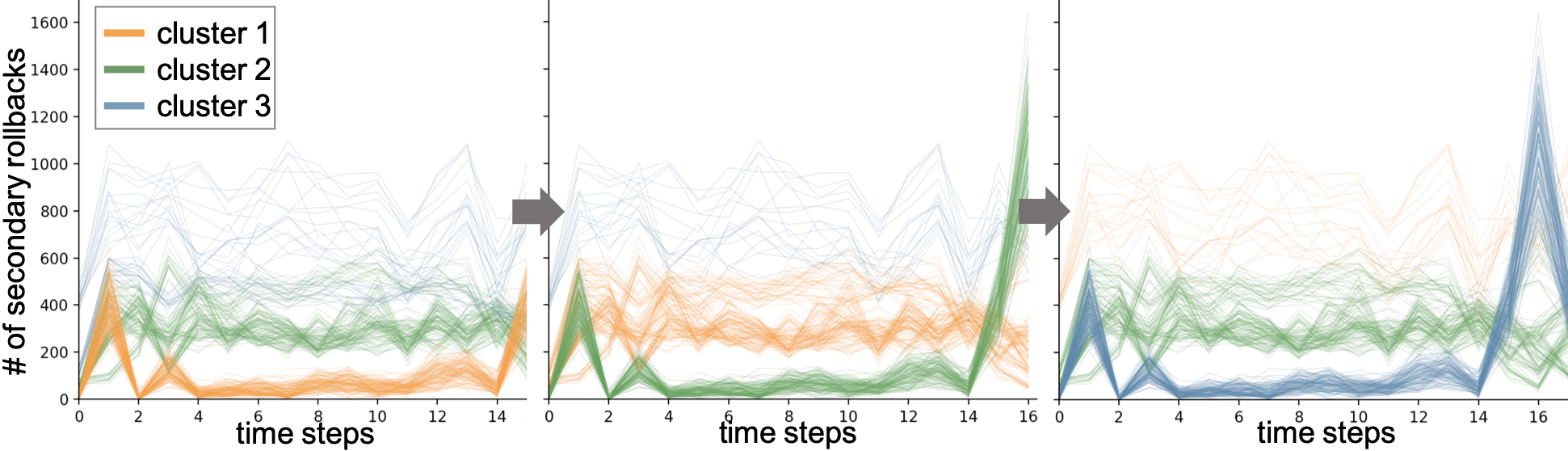}
     \label{fig:clustering_a}
    }
    \\
    \subfloat[With our cluster ID reassignment]{
     \includegraphics[width=0.95\linewidth,height=0.24\linewidth]{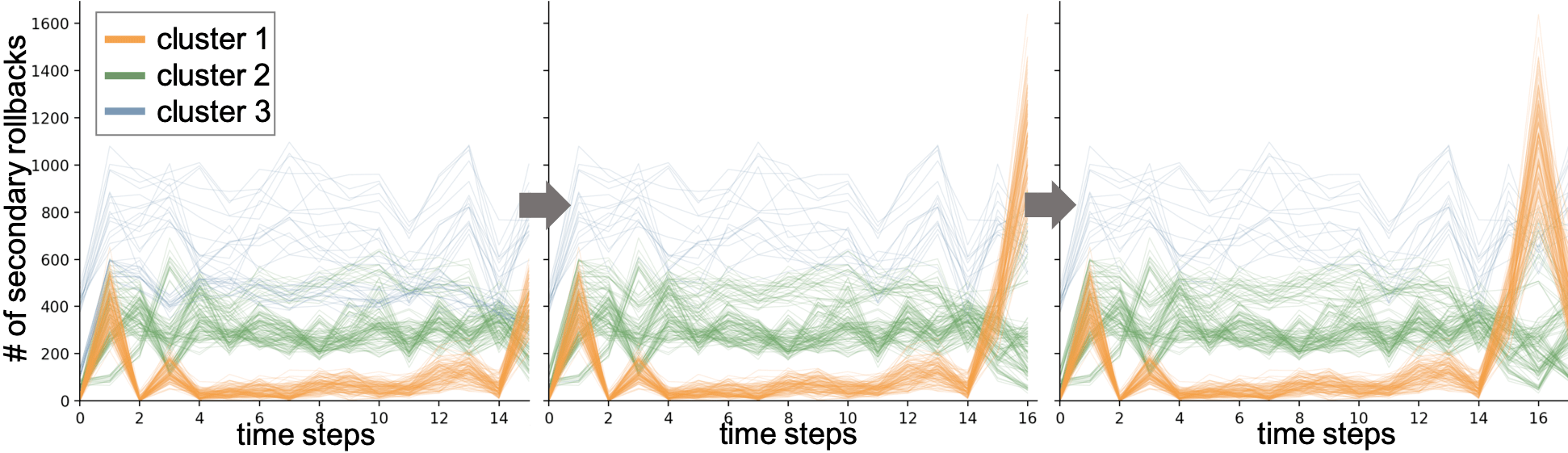}
     \label{fig:clustering_b}
    }
   	\caption{
   	The results of progressive time-series clustering (a) with and (b) without the cluster ID reassignments.
   	We plot the number of secondary callbacks for 256 entities with 15 (left), 16 (center), and 17 (right) time points.
   	We set \SI{50}{\milli s} as a required latency for this example and 147, 147, and 141 of 256 entities are processed to obtain the clusters with 15, 16, and 17 time points, respectively.
   	%The polyline colors represent the assigned cluster IDs with the method described in \autoref{sec:clutering}.
   	%The assigned cluster IDs in (a) drastically change in each result, whereas the results in (b) provide consistent cluster IDs.
   	}
	\label{fig:clustering}
\end{figure}

\vspace{-3pt}
\subsubsection{Progressive Time-Series Clustering}
\label{sec:clutering}
To support \textbf{R2}, we introduce a progressive time-series clustering method for streaming data where a fixed number of entities (e.g., routers) keep obtaining values (e.g., message packets sent) for each new time point (i.e.,  with the notations in \autoref{sec:data}, the fixed $n$ and the increasing $t$ for each metric). 
While many online clustering methods have been developed~\cite{bifet2018machine}, these methods are suited for streaming data only when a new entity will be fed with a fixed number of metrics (i.e., the increasing $n$ with the fixed $d$). 
% ~\cite{carnein2019optimizing,bifet2018machine}
% As described in \autoref{sec:data}, for performance data, we obtain a new feature (i.e., a value for a new time point) for all simulation entities.
Therefore, we cannot directly apply the existing methods to cluster entities' behaviors.

Instead, we develop a progressive clustering method for streaming data to obtain \del{the} reasonable results with low latency.   
Similar to \cite{turkay2017designing}, our method internally utilizes the incremental update mechanism used in online clustering. 
We employ mini-batch $k$-means clustering~\cite{sculley2010web} which is a variation of $k$-means clustering.
Unlike the ordinary $k$-means clustering, mini-batch $k$-means can incrementally update the clustering result with the new subset (or mini-batch) of entities.
When $n$ entities' values for a new time point arrive, we keep updating the clustering results with $m$ selected entities ($m \ll n$) until the specified latency or finishing to process all the data points.
% ($m \ll n$, $m$ must be larger than or equal to $k$ due to the limitation of mini-batch $k$-means clustering) 
Then, we obtain the final $k$ cluster centers and then assign each entity to a cluster that has the closest center. 
Since it might not be able to process all $n$ entities within the specified latency, the processing order would affect the clustering result. 
In order to select a wide variety of entities, our algorithm randomly selects $m$ entities from each of $k$-clusters which are calculated from the previous data points. 
\autoref{fig:clustering} shows the clustering results of our progressive clustering with a latency of \SI{50}{\milli s}. 
We can easily note that three clusters indicated with the orange, green, blue colors have different behaviors. 
% We can easily note that our clustering results produce similar results with the ordinary $k$-means clustering with a latency of less than 50 msec.

Another issue in applying a clustering method to streaming performance data is the consistency in the assigned cluster IDs between the previous and updated results. 
As shown in \autoref{fig:clustering_a}, most of the clustering methods including mini-batch $k$-means clustering generate the arbitrary order of cluster IDs for each execution. 
This causes a critical issue in the user's mental map preservation because the polyline colors would be changed every time when each entity obtains a value for a new time point. 
To solve this issue, our algorithm reassigns cluster IDs in the updated clustering results based on the relative frequency of previous cluster IDs which were assigned to each updated cluster ID's entities. 
Similar to the progressive clustering as described above,  we incrementally calculate the relative frequency by checking a randomly picked-out entity from each updated cluster ID to provide the result within the specified latency.
Afterward, our algorithm sorts the relative frequency among all the updated cluster IDs and then reassigns the updated cluster ID to the previous cluster ID from which has the highest relative frequency. 
\autoref{fig:clustering_a} and \ref{fig:clustering_b} show visualized results without and with our cluster ID reassignment, respectively.
We can see that our method reduces unnecessary color changes between the previous and updated results.

\begin{figure}[tb]
    \captionsetup{farskip=0pt}% <--- no gap at the top
	\centering
	\subfloat[Without the Procrustes transformation]{
     \includegraphics[width=0.8\linewidth,height=0.25\linewidth]{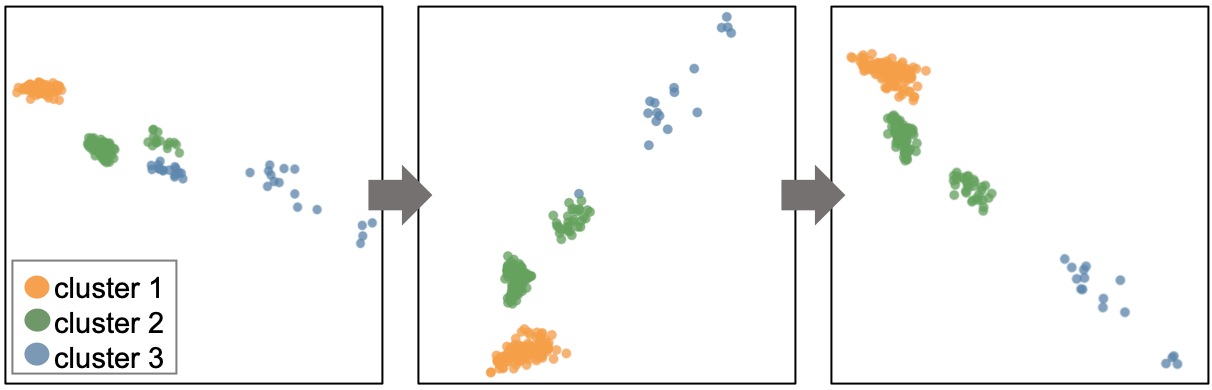}
     \label{fig:dr_a}
    }
    \\
    \subfloat[With the Procrustes transformation]{
     \includegraphics[width=0.8\linewidth,height=0.25\linewidth]{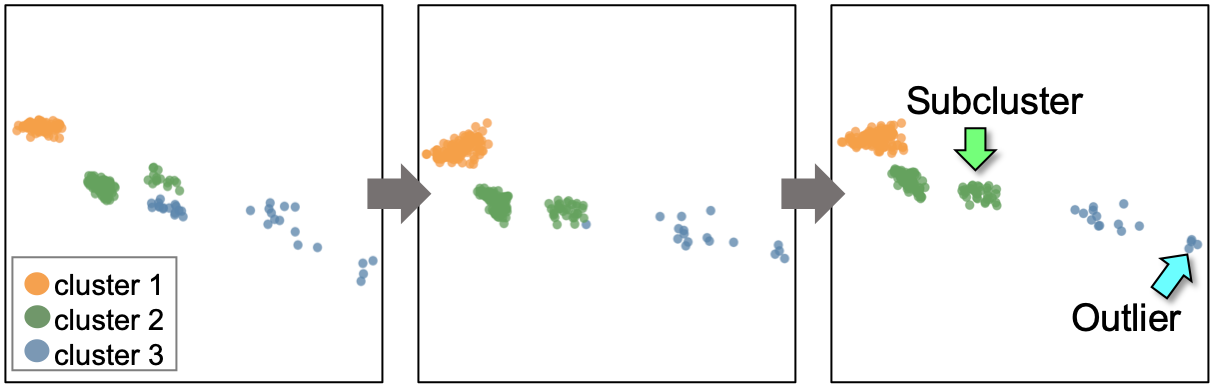}
     \label{fig:dr_b}
    }
   	\caption{
   	The results of progressive time-series DR (a) with and (b) without the Procrustes transformation. 
   	The same time-series and cluster IDs shown in \autoref{fig:clustering_b} are used.
   	In this example, we set \SI{1}{\milli s} as a required latency and 132, 130, and 138 of 256 entities are processed for the left, center, and right results, respectively.
   	%The point colors represent the cluster IDs assigned in \autoref{fig:clustering_b}.
   	}
	\label{fig:dr}
\end{figure}

\subsubsection{Progressive Time-Series Dimensionality Reduction}
\label{sec:dr}
To further support \textbf{R2}, we introduce progressive time-series dimensionality reduction (DR). 
DR methods can visualize (dis)similarities of temporal behaviors among entities as spatial proximities in a lower-dimensional plot (typically 2D). 
DR methods supplement analyzing temporal behaviors with clustering methods because DR results can help reveal small clusters (e.g., subclusters) and outliers (e.g., entities which have abnormal behaviors)\add{,} which are difficult to be found when using only clustering methods.
However, similar to the situation of time-series clustering, the existing online DR methods (e.g., \cite{ross2008incremental,pezzotti2017approximated,fujiwara2019incremental}) are not designed to update a lower-dimensional representation when new values arrive.

Therefore, we again adopt the incremental update used in incremental DR methods to obtain reasonable results within a required latency. 
We employ IPCA~\cite{ross2008incremental}. 
To obtain principal components (PCs), we keep updating the IPCA result with $m$ selected entities until the specified latency or finishing to process all the entities. 
% ($m$ must be larger than or equal to 2 due to the limitation of their IPCA)
Afterward, we project all entities' time-series into a 2D plot with the first and second PCs.
We use the same processing order as \autoref{sec:clutering}.
%As a result\add{, from \autoref{fig:dr_b},} \del{in \autoref{fig:dr_b} shows,with the DR result,} we can easily discern a subcluster and an outlier from the main clusters.
\autoref{fig:dr_b} shows results obtained with our method. 
From \autoref{fig:dr_b}, we can\,easily\,discern\,a\,subcluster\,and an\,outlier\,from\,the\,main\,clusters.

As discussed in \cite{fujiwara2019incremental}, (incremental) PCA also has an issue in the user's mental map preservation because PCA generates arbitrary sign flipping and rotations.
An example of this issue is shown in \autoref{fig:dr_a}.
We apply the Procrustes transformation~\cite{gower2004procrustes} used in \cite{fujiwara2019incremental} to provide consistent plots between previous and updated results. 
The Procrustes transformation tries to find the best overlap between two sets of positions (i.e., the previous and updated PCA results in our case) by using a combination of translation, uniform scaling, rotation, and reflection.
%(or reflection).
\autoref{fig:dr_b} shows visualized results with the Procrustes transformation.
In \autoref{fig:dr_b}, we can see that flips of data points' positions along $y$-direction in \autoref{fig:dr_a} are resolved. 

\subsubsection{Progressive Causal Relation Analysis Methods}
\label{sec:causality}
To support \textbf{R3}, we introduce progressive causal relation analysis methods. 
We support three methods: Granger causality test~\cite{hamilton1994time}, impulse response function (IR)~\cite{hamilton1994time}, and forecast error variance decomposition (VD)~\cite{hamilton1994time}. %~\cite{swanson1997impulse}
% ~\cite{granger1969investigating}
% Granger causality is defined ``in terms of predictability: a variable $X$ causes another variable $Y$, with respect to a given universe or information set that includes $X$ and $Y$, if present $Y$ can be better predicted by using past values of $X$ than by not doing so, all other information contained in the past of the universe being used in either case.''~\cite{pierce1977causality}.
Based on the definition of Granger causality, we can judge ``a time series ${X_t}$ `causes' another time series ${Y_t}$... if present $Y$ can be predicted better by using past values of $X$ than by not doing so''~\cite{pierce1977causality}.
Granger causality test evaluates a statistical hypothesis test where one time-series has this ``Granger causality'' from/to another time-series. 
Similar to the other statistical tests, Granger causality test provides whether there exists Granger causality between two time-series with $p$-value. 
However, with Granger causality test, we cannot measure how much one time-series affects another time-series. 
IR and VD can provide such quantitative information.
IR describes how much a shock to a variable of interest at a certain moment affects the other variables at subsequent time points. 
%An impulse-response function describes the evolution of the variable of interest along a specified time horizon after a shock in a given moment
%Impulse response functions show the effects of shocks on the adjustment path of the variables. 
On the other hand, VD provides the contribution of a shock to a variable of interest at a certain moment to the variance of the forecast error of other variables.
An example of the result of Granger causality test, IR, and VR can be found in \autoref{sec:causality_view}.
%A forecast error variance decomposition – or just variance decomposition for short – is a way to quantify how important each shock is in explaining the variation in each of the variables in the system.
%Forecast error variance decompositions measure the contribution of each type of shock to the forecast error variance. 
%It determines how much of the forecast error variance of each of the variables can be explained by exogenous shocks to the other variables.

All of these three analysis methods can be used after fitting one multivariate time-series to a vector autoregression (VAR) model~\cite{hamilton1994time}.
%(Some explanation for VAR?)
However, there are two challenges in applying VAR model fitting to streaming performance data.
One problem is that because HPC performance data consists of multiple entities, there are multiple time-series for each variable (i.e., metric), and thus, we cannot directly fit to a VAR model. 
To solve this problem, we use IPCA with the sign adjustment described in \autoref{sec:cpd}. 

Another problem is the high computational cost of VAR fitting: $O(t^2)$ ($t$ is the number of time points).
However, no incremental algorithm is available for VAR model fitting.
To provide a result within around a specified latency, we adaptively control the numbers of time points used in VAR model fitting. 
Our progressive VAR fitting starts with $s$ time points ($s \ll t$, we set $s=10$ as a default), which are randomly selected from $t$ time points.
We then obtain the result with $s$ time points in $t_c$ completion time.
Let $t_l$ be the user-specified latency.
Because the computational cost of VAR fit is $O(t^2)$, from the first calculation with $s$ time points, we can roughly estimate how many time points we can fit to a VAR model within remaining time $t_r$ (i.e., $t_r = t_l - t_c$).
This estimation calculated with $s \sqrt{t_r / t_c}$.
We update $s$ with $s \sqrt{t_r / t_c}$.
Then, the updated $s$ is used for the next VAR fitting.
These steps will be continued until $t_r \leq 0$ or obtaining a VAR model using all $t$ time points.

Also, when we obtain a new time point, we can expect that the number of time points which can be processed will be similar to the last $s$ in the previous progressive VAR fitting. 
Therefore, we start by using this $s$ in progressive VAR fitting when obtaining a new time point.
If the completion time $t_c$ in the previous VAR fitting with the last $s$ is larger than the required latency $t_l$, our method updates $s$ with $s \sqrt{t_l / t_c}$ and uses this updated $s$ for the next VAR fitting.

\subsubsection{Performance Evaluation}
\label{sec:perf_eval}
While the qualities of results with our methods depend on the existing methods used as the bases (e.g., IPCA), we evaluate computational performance for each method to provide reference information of the handleable data size.
We use an iMac (Retina 5K, 27-inch, Late 2014) with 4 GHz Intel Core i7, 16 GB 1,600 MHz DDR3.

For the online CPD method, the computational cost differs based on the number of time-series (i.e., $n$ entities) which we will reduce to the representative time-series with IPCA. 
Therefore, we measure the completion time for each different $n$ (from 100 to 100,000), as shown in Table\autoref{table:perf_eval_cpd}.
The computation of the progressive time-series clustering depends on the number of iterations in the $k$-means algorithm, the number of clusters to form, and the data length for each time-series (i.e., $t$ time points). 
We use fixed numbers of 100 and 3 for the numbers of iterations and clusters, respectively. 
Then, with different $t$ values from 100 to 100,000, we measure the number of time-series the algorithm processed from 10,000 time-series in one second. 
The result is shown in Table\autoref{table:perf_eval_clust}.
Similarly, as shown in Table\autoref{table:perf_eval_dr}, we evaluate the method in Table\autoref{sec:dr} with different $t$ values because its computation depends on the value of $t$.
Lastly, for the progressive VAR fitting, the number of time points can be processed is different based on the number of measured metrics (i.e., $d$ metrics). 
Thus, as shown in Table\autoref{table:perf_eval_var}, we measure the number of time points processed from 10,000 time points with multiple numbers of $d$.

From the results in \autoref{table:perf_eval}, we can see that the online CPD is fast even when $n$ is large (e.g., \SI{6.23}{\milli s} for $n=100,000$). 
With a setting of \SI{1}{s} latency, the clustering and DR methods processed large numbers of entities up until $t=10,000$.
This could be a reasonable amount of processed entities when we want to analyze several or tens of thousand entities. 
Also, the VAR fitting processed more than 800 time points even when $d=100$.
The quality of the results with the progressive algorithms depends on how much of the input data would be processed.
Thus, we should consider the balance of required latency and the size of input data based on available computational resources.
The performance results above can help us decide the granularity level of entities to be analyzed (e.g., compute node or CPU core level), the width of the time window, and the number of performance metrics collected. 

\begin{table}
    \renewcommand{\arraystretch}{0.6}
    \small
    \centering
    \captionsetup{farskip=0pt}% <--- no gap at the top
    \caption{The\,performance\,evaluation\,results\,((b)-(d):\,processed\,in\,\SI{1}{s}).
    %Each result is the average of ten executions.
    }
    \label{table:perf_eval}
    \subfloat[][Online CPD]{
        \label{table:perf_eval_cpd}
        \begin{tabular}{rr}
            \hline
            $n$ & completion time\\
            \hline
            100 & \SI{0.01}{\milli s}\\
            1,000 & \SI{0.05}{\milli s}\\
            10,000 & \SI{0.46}{\milli s}\\
            100,000 & \SI{6.23}{\milli s}\\
            \hline
        \end{tabular}
    }
    \quad
    \subfloat[][Progressive Clustering]{
        \label{table:perf_eval_clust}
        \begin{tabular}{rr}
            \hline
            $t$ & \ \ \ \  \# of entities \\
            % $t$ & processed in \SI{1}{s}\\
            \hline
            100 & 5,392\\
            1,000 & 4,921\\
            10,000 & 3,074\\
            100,000 & 562\\
            \hline
        \end{tabular}
    }\\
    \vspace{1pt}
    \subfloat[][Progressive DR]{
        \label{table:perf_eval_dr}
        \begin{tabular}{rr}
            \hline
            $t$ & \ \ \ \ \ \ \# of entities \\
            % $t$ & processed in \SI{1}{s}\\
            \hline
            100 & 10,000\\
            1,000 & 10,000\\
            10,000 & 2,396\\
            100,000 & 118\\
            \hline
        \end{tabular}
    }
    \quad
    \subfloat[][Progressive VAR fitting]{
        \label{table:perf_eval_var}
        \begin{tabular}{rr}
            \hline
            $d$ & \# of time points\\
            % $d$ & processed in \SI{1}{s}\\
            \hline
            10 & 10,000\\
            100 & 833\\
            \ \ 1,000 & 35\\
            \hline
        \end{tabular}
    }
\end{table}

%% file: 6_visualization_module.tex
\subsection{Interactive Visualization Module}
\label{sec:visualization}

\begin{figure*}[tb]
	\centering
	\includegraphics[width= 0.93\linewidth]{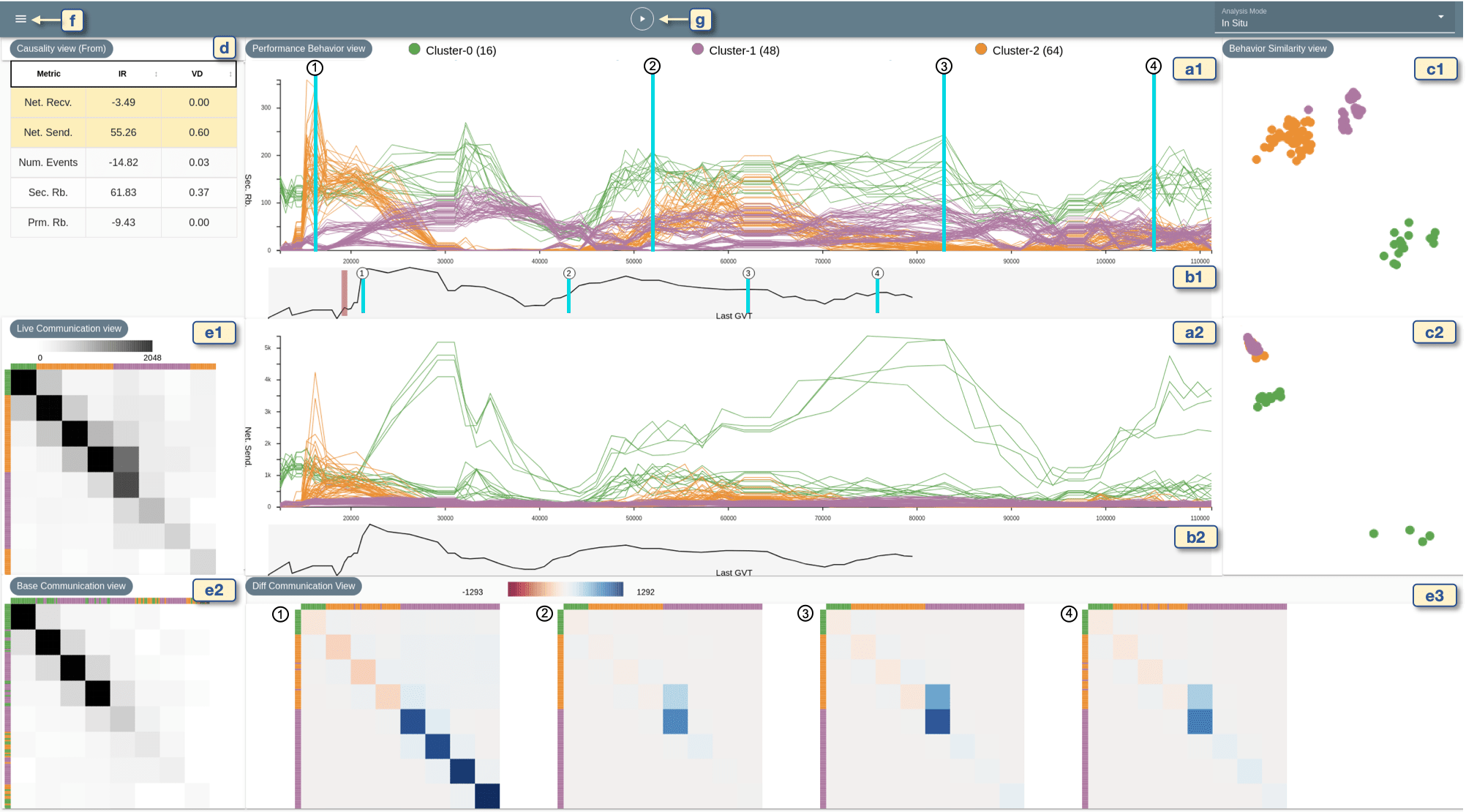}
  	\caption{
  	The user interface of our prototype system consisting of multiple views: (a1, a2) the performance behavior views, (b1, b2) histories of performance behaviors, (c1, c2) behavior similarity views, (d) metric causality view, and (e1, e2, e3) communication behavior views. (f) and (g) are the drop-down lists for settings and the button for pausing and resuming streaming updates, respectively.
  	}
	\label{fig:system_overview}
\end{figure*}

This module provides a user interface, as shown in \autoref{fig:system_overview}, which comprises multiple views to fulfill all the design requirements.
To better illustrate our visualizations and interactions, we provide a demonstration of the user interface as a supplementary video~\cite{supp}.

\subsubsection{Performance Behavior Views}
To fulfill \textbf{R1}, \textbf{R2}, and \textbf{R3}, the performance behavior views, as shown in \autoref{fig:system_overview}(a1, a2), provide visualizations with the change points (refer to \autoref{sec:cpd}) and clusters obtained (refer to \autoref{sec:clutering}) in the analysis module.
Using polylines, each performance behavior view presents the temporal changes of a performance metric for all the entities. 
Since the collected data comprises information from $n$ entities, each view shows $n$ polylines for each selected metric.
$x$- and $y$-coordinates represent a time point and a metric value respectively.
From the drop-down lists displayed by clicking the icon in the top left (\autoref{fig:system_overview}(f)), the analyst can select two metrics from $d$ collected metrics for the top (a1) and bottom (a2) views.
For example, in \autoref{fig:system_overview}, as indicated by the $y$-axis labels, `Sec.~Rb.' (i.e., Secondary Rollback) and `Net.~Send.' (i.e., Network sends) are selected.
We provide two views to support the comparison of performance behaviors for two different metrics, especially to review the causal relations (\textbf{R3}), as described in \autoref{sec:causality_view}. 

The online CPD described in \autoref{sec:cpd} is continuously applied to the time-series shown in the top view as a default. 
The analyst also can apply the CPD to the bottom view by selecting from the drop-down list in  \autoref{fig:system_overview}(f).
When a change point is detected, the corresponding time point (i.e., $x$-coordinate) is indicated with teal vertical lines in the corresponding view. 
For example, four change points are detected in \autoref{sec:causality_view}.

Similarly, the progressive clustering described in \autoref{sec:clutering} is applied to the top view as a default. 
The metric to perform clustering and the number of clusters can be modified from the drop-down lists in \autoref{fig:system_overview}(f). 
Obtained cluster IDs are encoded as the colors of polylines.
The categorical colors are chosen to provide sufficient contrast for distinguishing the clusters. 
A legend for the cluster IDs are shown in the top of \autoref{fig:system_overview}(a1). 
The number in each parenthesis represents the number of entities belonging to each cluster (e.g., Cluster-0 contains 16 entities). 
Also, to make the comparison between two different metrics easier using the top and bottom views, the same colors that represent the cluster IDs are used for the corresponding entities. 
For example, in \autoref{fig:system_overview}, the bottom view is colored based on the cluster IDs computed for the top view. 
We can see that green lines (i.e., Cluster-0) in both top and bottom views have high `Sec.~Rb.' and 'Net.~Send.' values in many time points.

Since the data stream continuously feeds data for a new time point, visualizing all the fed time points consumes the visual space and, as a result, it becomes challenging to covey the detailed changes.  
Therefore, as new data arrive, we slide each view's time window for the visualization. 
Also, to provide the historical context with the past time points from the start of the simulation, as shown in \autoref{fig:system_overview}(b1, b2), the summary behavior placed at the bottom of each view shows the average value across entities for each time point.

\subsubsection{Behavior Similarity Views}
As discussed in \autoref{sec:dr}, providing DR results along with cluster IDs can supplement analyzing temporal behaviors, such as finding outliers (\textbf{R2}). 
We apply the method described in \autoref{sec:dr} to multiple time-series shown in each performance behavior view (\autoref{fig:system_overview}(a1) and (a2)).
Then, the behavior similarity views visualize the corresponding DR results (i.e., \autoref{fig:system_overview}(c1) and (c2) show the result from (a1) and (a2), respectively).
In addition, to easily relate with the clustering results, we color the points with the corresponding colors used for encoding the cluster IDs in the performance behavior views. 
For example, all polylines and points shown in (a1, a2, c1, c2) are colored based on the clustering IDs obtained in (a1).
From (c1), We can see that each entity's behavior of `Sec.~Rb.' is clearly separated into three clusters.
However, in (c2), Cluster-0 (green) has four outliers as shown around the center. 
This can further be validated using the bottom behavior view (a2). 
The four green lines have much higher values when compared with the other green lines. 

\subsubsection{Metric Causality View}
\label{sec:causality_view}
The metric causality view (\autoref{fig:system_overview}(d)) provides the results from the causality analysis described in \autoref{sec:causality} between the performance metrics (\textbf{R3}).
This view can help the analyst determine which metrics share causal relationships with the chosen metric of interest for the top performance behavior view (\autoref{fig:system_overview}(a1)).
Using Granger causality, we can derive two kinds of causal relationships: (1) from-causality: effects from other metrics on the metric of interest, and (2) to-causality: effect of metric of interest on other metrics. 
We display the from-causality results by default because the analyst often wants to review a metric which conveys the performance decrements first and then to identify which metrics likely cause it.
However, the analyst can change to the to-causality results from the settings in \autoref{fig:system_overview}(f). 
We inform metrics of which $p$-value for Granger causality test is less than the user-defined value (the default is $0.05$) with the yellow background.
Also, we display the results of IR and VD. 
The metrics can be sorted based on the values for IR and VD by clicking the column title.
Causality analysis results are useful for the analyst to decide the second metric to be shown in the bottom performance behavior view (\autoref{fig:system_overview}(a2)).
For example, in~\autoref{fig:system_overview}, because we can see that `Net.~Send' has Granger causality and the highest IR value, we select and visualize its behavior in (a2).

\subsubsection{Communication Behavior Views}
\label{sec:comm_behavior_view}
The communication behavior views shown in \autoref{fig:system_overview}(e1, e2, e3) visualize the communication patterns between the entities (\textbf{R4}).
We use an adjacency-matrix based visualization to show communications which can be represented as a weighted graph (refer to \autoref{sec:data}).
Using the adjacency-matrix can provide flexibility to support any type of network topologies.
While reviewing communications at the latest time point is not enough to grasp the changes in communication patterns, looking through communications at all the time points is not realistic in the streaming setting. 
Therefore, we support three different views to provide a summary of communication patterns and their changes. 
The details of each view are described below.

% \vspace{0pt}
\noindent\textbf{Live communication view (\autoref{fig:system_overview}(e1))} visualizes communications among entities during the sampling interval at the latest time point.
While each row and each column of the matrix corresponds to one entity with the user-defined order, each cell's color represents the number of communications between the corresponding row and column.
We use a gray sequential colormap (darker gray denotes higher communication).
Additionally, to inform the cluster information of each entity, the cluster ID corresponding to each row/column is encoded as colored rectangles on the top and left sides of the adjacency matrix.
For example, in \autoref{fig:system_overview}(e1), the entities in the green cluster placed at the first several rows and columns cause many communications within the entities in the same cluster.

To provide better scalability, this view can cooperate with the hierarchical information of entities, which often can be seen in HPC systems. 
For example, many HPC systems consist of multiple racks, compute nodes, CPUs, and cores.
These have hierarchical relationships (e.g., a compute node contains multiple CPUs). 
The analyst can choose the granularity of entities (i.e., the level of hierarchy) to be shown in this view. 
When the lower granularity is selected instead of the original granularity, each cell shows the average amount of communications within and among the groups of entities. 
For example, in \autoref{fig:system_overview}(e1), because many entities (128 entities) will make the available space for each cell small, we select one lower granularity consisting of 8 groups (i.e., each group contains 16 entities). 
This view also allows the analyst to show the communications with one higher granularity level by double-clicking the grouped cell.

\vspace{0pt}
\noindent\textbf{Base communication view (\autoref{fig:system_overview}(e2))} visualizes the communication matrix at the user-selected time point. 
We use the same visualization methods as for the live communication view.
The analyst can select the time point using a brown draggable vertical line placed on the summary behavior view. 

\noindent\textbf{Diff communication view (\autoref{fig:system_overview}(e3))} shows communications at multiple time points which are essential to understand the changes in communications. 
We use the change points identified by CPD as such essential time points because each change point shows significant changes from the previous time points. 
We order each matrix corresponding to each change point along the horizontal direction from left to right. 
Also, we indicate the corresponding change point with the numerical labels placed in \autoref{fig:system_overview}(a1, b1, e3).

To allow the analyst to compare the communications at each change point with the one at the selected time point for the base communication view (\autoref{fig:system_overview}(e2)), each matrix's cell shows the difference in the amount of communication between the selected time and each change point.
As shown in the colormap placed in \autoref{fig:system_overview}(e3), the differences are encoded using a red-blue divergent colormap where the darker red and blue represent higher positive and negative values, respectively. 
The comparison with the selected base communications provides the flexibility in the analysis. 
For example, when the analyst selects a time point which has no communications (e.g., the start time point), the diff communication view visualizes the amounts of communications as they are. 
On the other hand, when the analyst selects one of the change points as a base, the view shows how the communications are changed from the selected change point. 
This helps understand which entities' communications affect the changes in the metric visualized in the performance behavior views. 

\subsubsection{User Interactions across Views}
\label{sec:user-interactions}
In addition to the interactions in each view, we provide several interactions that are linked to multiple views.

% \vspace{0pt}
\noindent\textbf{Pausing and resuming streaming updates.}
Since the visualization module updates its views as the analysis module sends new results, the updates occur at data collection rate.
To provide a mechanism that allows the analyst to interact with the visualized results even when the data collection rate is high, we provide a \textit{pause} button (\autoref{fig:system_overview}(g)) to pause the views from updating. 
Also, after pausing, the button is toggled to a \textit{resume} button. 
When the \textit{resume} button is clicked, the views return back to the current state of the simulation. 

\noindent\textbf{Selecting a time point of interest.}
As described in \autoref{sec:comm_behavior_view}, the analyst can select a time point of interest in the summary of performance behavior shown in \autoref{fig:system_overview}(b1). 
Selecting a time point immediately updates the base and diff communication views. 

\noindent\textbf{Selecting entities of interest.}
Although a set of views can reveal overall patterns in performance and communication behaviors, the user would be interested in to analyze a subset of entities (e.g., a subcluster appeared in the behavior similarity views) in more detail. 
Therefore, we provide fundamental linking-and-brushing interactions. 
For example, the behavior similarity views support lasso selection to choose entities of interest. 
Once selected, the polylines in the performance behavior views belonging to the selected entities are highlighted by coloring the rest with gray and low opacity.

%% file: 7_case_studies.tex
\section{Case Studies}
\label{sec:case_study}
For demonstrating the applicability and effectiveness of our framework, we analyze the streaming data from parallel discrete-event simulation (PDES).
We show that our visual analytics system can be used to effectively identify performance problems, investigate the source of the problem, and develop insights for improving performance.
Besides, we collected feedback from the domain experts, which confirmed the usefulness of our framework and provided further design considerations.
PDES is used for studying complicated scientific phenomenon and systems.
PDES typically runs on HPC systems for a long period using many compute nodes, and thus, simulations with low efficiency cost significant energy and time.
Therefore, it is crucial to optimize performance and relieve bottlenecks.
For distributed and parallel computing, PDES distributes a group of processes, called processing elements (PEs), that run across compute nodes.
PEs communicate by exchanging time-stamped event messages that are processed to ensure the correct order of events (i.e., the future event must not affect the past event)~\cite{fujimoto1990parallel}. 
As PDES, we use the Rensselaer's Optimistic Simulation System (ROSS)~\cite{carothers2002ross}, an open-source discrete-event simulator.
ROSS uses optimistic parallel event scheduling, where each PE contains a group of logical processes (LPs) that can independently process events to avoid frequent global synchronization with other PEs.
Also, ROSS introduces kernel processes (KPs) for managing a shared processed event list for a collection of LPs mapped to a single PE, which has been demonstrated to process up to billions of events~\cite{mubarak2012modeling,bauer2009scalable}.
When a KP rolls back, it must roll back all its LPs to the same point in virtual time.
Therefore, having too many LPs mapped to a KP can result in performance degradation due to unnecessary rollbacks of other LPs in the same KP.
Furthermore, the performance of ROSS depends on the model being simulated and the associated workload, which makes it difficult to investigate performance bottlenecks.

\subsection{Streaming PDES Data}
\label{sec:case_study_data}
In ROSS, MPI is used for distributed and parallel computing, where each PE is an MPI process, and message passing between PEs is performed asynchronously.
ROSS uses the global virtual time (GVT) for tracking the time in PDES, where the GVT is computed as the minimum of all unprocessed events across all PEs in the simulation~\cite{jefferson1985virtual}.
The streaming data collected from ROSS includes numerous metrics at the PE, KP, and/or LP levels. 
In particular, we focus on the metrics that relate to rollback behaviors (e.g., the number of rollbacks) or communications among PEs, KPs, and LPs (e.g., the number of events sent or received), which can significantly affect the efficiency of PDES. 
In the case studies, we analyze the performance by comparing the number of rollbacks during the GVT computation. 
To perform this, we use our data management module to collect the following metrics just after the GVT computation occurs (`Last GVT'). 

\begin{compactitem}
    \item \textbf{Primary Rollbacks (Prim.~Rb.)}: The number of rollbacks on a KP caused by receiving an event out of order.
    \item \textbf{Secondary Rollbacks (Sec.~Rb.)}: The number of rollbacks on a KP caused by an anti-message (i.e., a cancellation message).
    \item \textbf{Network Sends (Net.~Send.)}: The number of events sent by LPs over the network.
    \item \textbf{Network Receives (Net.~Recv.)}: The number of events received by LPs over the network.
    \item \textbf{Number of processed events (Num.~Events)}: The total number of processed events by LPs over the network.
\end{compactitem}

We set up the ROSS with the Dragonfly network~\cite{dragonfly} simulation model provided by the CODES~\cite{mubarak2012modeling}.
The simulated Dragonfly network is similar to the network used by Theta Cray XC supercomputer at Argonne National Laboratory with 864 routers.
The simulations are run with 8 to 16 PEs, where each PE has 16 KPs, with up to 16,384 LPs.

% \subsection{Real-Time Analysis of PDES Performance}
\subsection{Monitoring Key Changes in PDES Performance}
Because PDES uses a large number of entities in a complex structure, identifying the causes and sources of performance problems is difficult.
With our visual analytics framework, we can effectively monitor the changes in the key metrics in real time, and analyze which computing entity is causing performance problems. 

Since the efficiency of optimistic PDES depends on the number of rollbacks (including both primary and secondary rollbacks), we start to monitor each KP's `Sec.~Rb.' and `Prim.~Rb.' with the performance behavior views. 
We set the number of clusters to be 3, and the cluster IDs are calculated based on `Sec.~Rb.'. 
From monitoring the causality results, as shown in \autoref{fig:system_overview}(d), we notice that `Net.~Send.' has the Granger causality to `Sec.~Rb.' and has a large IR value. 
Hence, we update our second metric to `Net.~Send' and continue to study the influence of `Net.~Send' on `Sec.~Rb.'.

After the online CPD proceeded with several GVT intervals, we can see the first change point \textcircled{\small 1} (refer to \autoref{fig:system_overview}(a1)). 
\textcircled{\small 1} clearly shows that  the ``Sec.~Rb.'' has drastically increased (Cluster-2 with the orange-colored polylines).
This alerts us that we should closely monitor the rollback behaviors because the efficiency of the simulation can be significantly decreased if the number of secondary rollbacks stays high.
\autoref{fig:system_overview}(a1) and (a2) show the result after the simulation is run for 75 GVT intervals.
The online CPD detected four change points (\textcircled{\small 1}, \textcircled{\small 2}, \textcircled{\small 3}, and \textcircled{\small 4}).  
While the metric causality view shows the detailed behaviors of each entity, the summary behavior view shows the trends of the PDES as a whole.
As shown in (\autoref{fig:system_overview}(b1)), the average of secondary rollbacks have a peak at the start of the simulation close to \textcircled{\small 1}, then reduces between \textcircled{\small 1} and \textcircled{\small 2}, and increases again from \textcircled{\small 2} to \textcircled{\small 3}, and maintains the values till \textcircled{\small 4}.
The summary view with CPD allows us to confirm the existence of performance problems in the simulations as the secondary rollbacks periodically increase at different points.

\begin{figure}[tb]
	\centering
	\includegraphics[width=1.0\linewidth]{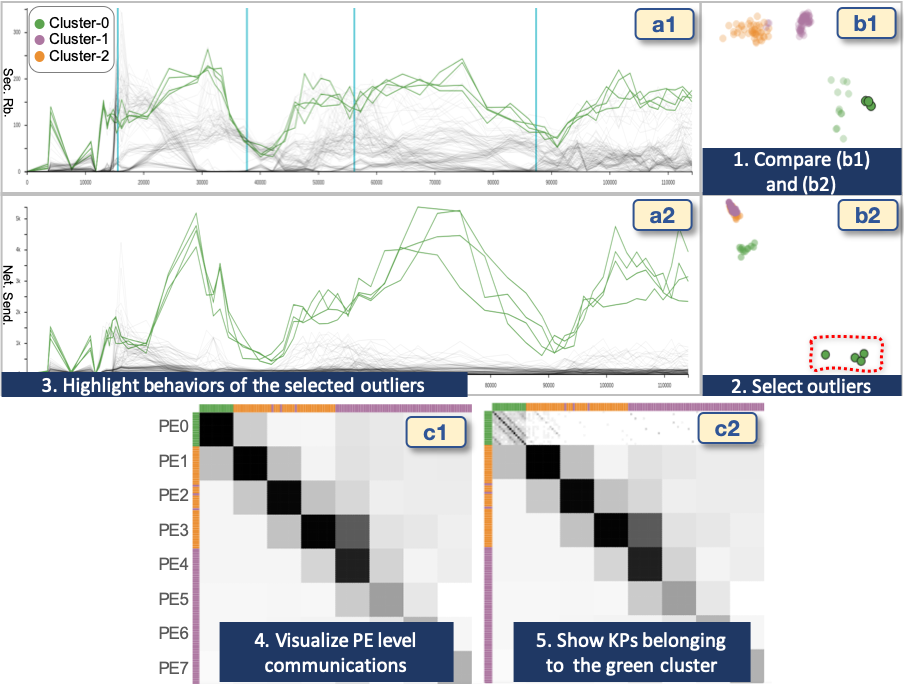}
  	\caption{Detailed analysis of performance bottlenecks.}
	\label{fig:case-study-1}
\end{figure}

\subsection{Tracing Performance Bottlenecks}
\label{sec:case_subset}
If we can confirm the existence of performance problems, we can stop the simulation to prevent wasting time and energy by running the long and problematic PDES.
Thus, we move on to the identification of the source of the problems, and our system can be used to trace the performance issues and bottlenecks.
As shown in \autoref{fig:system_overview}(a1, a2), we can see that some of green-colored polylines (i.e., Cluster-0) continue to show relatively high values in both `Sec.~Rb.' and `Net.~Sends'. 
Thus, we further investigate the green cluster.
By associating the behavior similarity view (\autoref{fig:system_overview}(c1)) to the performance behavior view (\autoref{fig:system_overview}(a1)), we realize that all the KPs belonging to the green cluster have a high similarity with each other for their rollback behavior. 
However, as shown in \autoref{fig:system_overview}(c2), for their `Net.~Send.', they are separated into one major cluster at the top left and one small cluster that has four KPs at the right bottom.
To further analyze their behaviors, we select the four entities with the lasso selection. 
The result is shown in \autoref{fig:case-study-1}.
From the performance behavior views (a1, a2), we can discern that these four KPs cause large numbers of network sends and, as an influence, they also cause large numbers of secondary rollbacks.

\vspace{-1pt}
\subsection{Analyzing Communication Patterns}
After identifying the source of the performance problem, insights for removing the bottlenecks and optimizing the performance can be extracted from our visual analysis results.
Since a LP that communicates with other LPs that have high number of rollbacks can increase its own chances to have rollbacks, a better control on the communication events can reduce the number of rollbacks and improve efficiency.
Therefore, we analyze the rollback behaviors with the communication patterns in the live communication view, as shown in \autoref{fig:case-study-1}(c1, c2).
(\autoref{fig:case-study-1}(c1)) shows the communication between PEs or KPs with arrangements in rows and columns of the matrix based on their ranks.
At default, the communications between the PEs are shown.
We can see the cluster IDs of the KPs shown at the top and left of the matrix indicate that all the KPs in the green cluster (including the four KPs) belong to PE0. 
Also, we can see that KPs in the green cluster dominantly communicate with KPs within the same cluster.
To further drill down to the KP level, we click on the matrix cell belonging to PE0 (\autoref{fig:case-study-1}(c2)) and identify that all KPs, including the four KPs, generate many communications only within themselves (e.g., KP0 communicates to KP0).
This indicates that LPs managed by these KPs have high communications with each other.
Since these KPs has high number of rollbacks, they may cause other KPs communicating with them to have higher chances to have rollbacks.
To avoid these rollbacks, the mapping from LPs to KPs should be changed to alleviate the unbalanced communications.   

The case studies above show that our visual analytics framework can tackle the challenges of real-time performance monitoring and analysis of PDES.
With effective supports for analyzing the streaming PDES data in real-time, our framework helps the analyst identify the time points and locations (e.g., PEs and KPs) that performance issues occurred.
Once such a performance issue is realized, the analyst can stop the simulation to save energy and time.
The analysis results also provide hints to the cause of the performance issues, allowing PDES developers to debug or optimize performance.

\vspace{-1pt}
\subsection{Expert Feedback}
Our team consists of both visualization and HPC researchers that participated in the design and development of our framework, and we have conducted a preliminary evaluation of the visual analytics system with PDES developers and experts.
While the feedback confirmed the usefulness of our visual analytics framework and system for analyzing streaming PDES data, it also provided insights to improve the system further.
For identifying performance bottlenecks, the experts pointed out that providing multiple communication views with comparison to a reference point based on change point detection is useful, and suggested that allowing filtering based on the types or attributes of the communication can further help analyze performance problems.
For gaining insights to improve performance, the HPC researchers pointed out that allowing selection of outliers based on temporal behavior is useful as a first step, and suggested that our system could be more helpful if it can provide a mapping between the communication and performance metrics based on entities (e.g., KPs or MPI ranks).
Based on these comments, we plan to further develop and extend our framework in the future, as well as conducting a user study to understand better how users perform analysis on streaming data using our extended framework.

%% file: 8_discussion.tex
\section{Discussion and Limitations}
\vspace{0pt}
\textbf{Generality of our framework.}
As described in \autoref{sec:bg_and_dr} and \autoref{sec:framework}, our framework is designed to be able to handle streaming performance data, which contains multivariate time-series, communication data, or both. 
Additionally, our analysis algorithms and most of the visualizations are applicable to any streaming multivariate data.
Currently, streaming data becomes increasingly available  because of the emerging of the Internet of things. 
As an example, our algorithms can be used to analyze sensor data from electronic equipment in a building, such as vending machines and air conditioners.

\vspace{1pt}
\noindent\textbf{Algorithm selection.}
While our visual analytics framework leverages the existing incremental algorithms (e.g., incremental PCA), the analysis module is flexible enough to be replaced with improved algorithms based on the analysis needs. 
For example, while we have chosen the mini-batch $k$-means clustering~\cite{sculley2010web} because of its popularity, we can use the other clustering methods, such as the incremental version of density-based clustering~\cite{chen2002incremental}. 
In addition, the algorithms that we selected can provide visual consistency (i.e., the cluster ID reassignment and the Procrustes transformation) can also be integrated with any other clustering or DR algorithms~\cite{fujiwara2019incremental}. 

\vspace{1pt}
\noindent\textbf{Visualization selection.}
Employing complex and advanced visualizations for analyzing streaming data is hard for the user to track.
Thus, the HPC experts prefer intuitive and straightforward visualizations such as line charts and scatterplots.
Our framework also integrates techniques that allow the user to preserve the mental map and reveal trends within simulation performance to understand their behavior.
Although other visual techniques such as parallel coordinates could be employed to summarize the entity information, they would suffer scalability for a higher number of data dimensions and they are also difficult to convey temporal changes.
For the communication view, we initially considered using a hierarchical radial layout, as similar to Li et. al~\cite{li2017visual}, where entities are arranged along the radial axis and the links between the entities represented the communication.
The radial layout presents advantages in encoding multiple attributes along the outer radius and coupling with edge-bundling to summarize the communication.
However, the radial layout would pose visual challenges in a progressive setting as overlapping edges can affect the mental map, and communicating the differences can impose further design challenges.

\vspace{1pt}
\noindent\textbf{Limitations.}
Our framework has several limitations that need to be addressed in future work.
First, when performance data is collected in an extremely short sampling rate, the progressive algorithms could not provide useful intermediate results because the algorithms may process only a limited number of entities. 
Also, visualizations keep updating too frequently, and, as a result, the analyst may not be able to follow the changes. 
We can solve these problems by controlling the frequency of updates in the data management module.
For example, the data management module can feed the data to the other modules only when the changes between the current and past data are large. 
The second problem is the number of collected metrics is large (e.g., more than 20 metrics).
Even though our causality analysis methods are helpful in narrowing down the metrics that should be compared, many metrics might have causal relationships with each other. 
In this case, aggregating multiple metrics based on their similarities could be useful to obtain a smaller number of metrics. 
When the data contains an extremely-large number of entities (e.g., more than 10,000), the visualizations could be too cluttered. 
To ensure the perceptual scalability, the entities can be aggregated based on their similarities.
Also, the aggregation could be useful to depict a high-level overview of the data. 
However, we want to note that, for the streaming setting, performing the aggregation is not trivial. 
Because applying the aggregation for each update could frequently and drastically change the visualized results (e.g., the aggregated line shapes).
We would like to address this problem as our future work.

%% file: 9_conclusions.tex
\section{Conclusions}
Our framework contributes to gaining insights from streaming performance data in real-time with algorithmic and visual analytics supports. 
Real-time analysis of a large-scale simulation becomes more necessary as the computational capacity of HPC systems continues to grow.
Therefore, our work helps enhance the efficient utilization of advanced HPC systems hereafter.

%% file: 00_main.bbl
\begin{thebibliography}{10}

\bibitem{supp}
The supplementary materials.
\newblock \url{https://jarusified.github.io/2020-streaming.html}.

\bibitem{aggarwal2013survey}
C.~C. Aggarwal.
\newblock A survey of stream clustering algorithms.
\newblock In {\em Data Clustering}, pp. 231--258. Chapman and Hall/CRC, 2013.

\bibitem{albers2003online}
S.~Albers.
\newblock Online algorithms: a survey.
\newblock {\em Mathematical Programming}, 97(1-2):3--26, 2003.

\bibitem{bauer2009scalable}
D.~W. Bauer~Jr, C.~D. Carothers, and A.~Holder.
\newblock Scalable time warp on blue gene supercomputers.
\newblock In {\em Proc. ACM/IEEE/SCS Workshop on Principles of Advanced and
  Distributed Simulation}, pp. 35--44, 2009.

\bibitem{bhatele2016analyzing}
A.~Bhatele, N.~Jain, Y.~Livnat, V.~Pascucci, and P.-T. Bremer.
\newblock Analyzing network health and congestion in dragonfly-based
  supercomputers.
\newblock In {\em Proc. IEEE IPDPS}, pp. 93--102, 2016.

\bibitem{biddiscombe2007time}
J.~Biddiscombe, B.~Geveci, K.~Martin, K.~Moreland, and D.~Thompson.
\newblock Time dependent processing in a parallel pipeline architecture.
\newblock {\em IEEE TVCG}, 13(6):1376--1383, 2007.

\bibitem{bifet2018machine}
A.~Bifet, R.~Gavald{\`a}, G.~Holmes, and B.~Pfahringer.
\newblock {\em Machine learning for data streams: with practical examples in
  MOA}.
\newblock MIT Press, 2018.

\bibitem{bodenham2017continuous}
D.~A. Bodenham and N.~M. Adams.
\newblock Continuous monitoring for changepoints in data streams using adaptive
  estimation.
\newblock {\em Statistics and Computing}, 27(5):1257--1270, 2017.

\bibitem{carothers2002ross}
C.~D. Carothers, D.~Bauer, and S.~Pearce.
\newblock {ROSS}: A high-performance, low-memory, modular time warp system.
\newblock {\em Journal of Parallel and Distributed Computing},
  62(11):1648--1669, 2002.

\bibitem{chen2002incremental}
N.~Chen, A.-z. Chen, and L.-x. Zhou.
\newblock An incremental grid density-based clustering algorithm.
\newblock {\em Journal of software}, 13(1):1--7, 2002.

\bibitem{cheng2016framework}
S.~Cheng, K.~Mueller, and W.~Xu.
\newblock A framework to visualize temporal behavioral relationships in
  streaming multivariate data.
\newblock In {\em Proc. New York Scientific Data Summit}, pp. 1--10. IEEE,
  2016.

\bibitem{dasgupta2018human}
A.~Dasgupta, D.~L. Arendt, L.~R. Franklin, P.~C. Wong, and K.~A. Cook.
\newblock Human factors in streaming data analysis: Challenges and
  opportunities for information visualization.
\newblock {\em CGF}, 37(1):254--272, 2018.

\bibitem{de2013visualizing}
W.~De~Pauw, J.~Wolf, and A.~Balmin.
\newblock Visualizing jobs with shared resources in distributed environments.
\newblock In {\em Proc. IEEE Working Conf. Software Visualization}, pp. 1--10,
  2013.

\bibitem{fujimoto1990parallel}
R.~M. Fujimoto.
\newblock Parallel discrete event simulation.
\newblock {\em Communications of the ACM}, 33(10):30--53, 1990.

\bibitem{fujiwara2019incremental}
T.~Fujiwara, J.-K. Chou, S.~Shilpika, P.~Xu, L.~Ren, and K.-L. Ma.
\newblock An incremental dimensionality reduction method for visualizing
  streaming multidimensional data.
\newblock {\em IEEE TVCG}, 2019.

\bibitem{fujiwara2018visual}
T.~Fujiwara, J.~K. Li, M.~Mubarak, C.~Ross, C.~D. Carothers, R.~B. Ross, and
  K.-L. Ma.
\newblock A visual analytics system for optimizing the performance of
  large-scale networks in supercomputing systems.
\newblock {\em Visual Informatics}, 2(1):98--110, 2018.

\bibitem{fujiwara2017visual}
T.~Fujiwara, P.~Malakar, K.~Reda, V.~Vishwanath, M.~E. Papka, and K.-L. Ma.
\newblock A visual analytics system for optimizing communications in massively
  parallel applications.
\newblock In {\em Proc. IEEE VAST}, pp. 59--70, 2017.

\bibitem{gower2004procrustes}
J.~C. Gower, G.~B. Dijksterhuis, et~al.
\newblock {\em Procrustes problems}, vol.~30.
\newblock Oxford University Press on Demand, 2004.

\bibitem{hamilton1994time}
J.~D. Hamilton.
\newblock {\em Time series analysis}, vol.~2.
\newblock Princeton University Press, 1994.

\bibitem{isaacs2014combing}
K.~E. Isaacs, P.-T. Bremer, I.~Jusufi, T.~Gamblin, A.~Bhatele, M.~Schulz, and
  B.~Hamann.
\newblock Combing the communication hairball: Visualizing parallel execution
  traces using logical time.
\newblock {\em IEEE TVCG}, 20(12):2349--2358, 2014.

\bibitem{isaacs2014state}
K.~E. Isaacs, A.~Gim{\'e}nez, I.~Jusufi, T.~Gamblin, A.~Bhatele, M.~Schulz,
  et~al.
\newblock State of the art of performance visualization.
\newblock In {\em Proc. EuroVis (STARs)}, pp. 141--160, 2014.

\bibitem{jefferson1985virtual}
D.~R. Jefferson.
\newblock Virtual time.
\newblock {\em ACM Trans. Programming Languages and Systems}, 7(3):404--425,
  1985.

\bibitem{jolliffe1986principal}
I.~T. Jolliffe.
\newblock {\em Principal Component Analysis and Factor Analysis}, pp. 115--128.
\newblock Springer, 1986.

\bibitem{dragonfly}
J.~Kim, W.~J. Dally, S.~Scott, and D.~Abts.
\newblock {Technology-Driven, Highly-Scalable Dragonfly Topology}.
\newblock In {\em Proc. Int. Symp. Computer Architecture}, pp. 77--88, 2008.

\bibitem{krstajic2013visualization}
M.~Krstajic and D.~A. Keim.
\newblock Visualization of streaming data: Observing change and context in
  information visualization techniques.
\newblock In {\em Proc. IEEE Int. Conf. Big Data}, pp. 41--47, 2013.

\bibitem{landge2012visualizing}
A.~G. Landge, J.~A. Levine, A.~Bhatele, K.~E. Isaacs, T.~Gamblin, M.~Schulz,
  et~al.
\newblock Visualizing network traffic to understand the performance of
  massively parallel simulations.
\newblock {\em IEEE TVCG}, 18(12):2467--2476, 2012.

\bibitem{Larsen:ISAV17}
M.~Larsen, J.~Ahrens, U.~Ayachit, E.~Brugger, H.~Childs, B.~Geveci, and
  C.~Harrison.
\newblock {The ALPINE In Situ Infrastructure: Ascending from the Ashes of
  Strawman}.
\newblock In {\em Proc. ISAV 2017}, pp. 42--46.

\bibitem{li2019visual}
J.~K. Li, T.~Fujiwara, S.~P.~Kesavan, C.~Ross, , C.~D. Mubarak,
  Misbah~Carothers, R.~Ross, and K.-L. Ma.
\newblock A visual analytics framework for analyzing parallel and distributed
  computing applications.
\newblock In {\em Proc. Symp. Visualization in Data Science}. IEEE, 2019.

\bibitem{li2017visual}
J.~K. Li, M.~Mubarak, R.~B. Ross, C.~D. Carothers, and K.-L. Ma.
\newblock Visual analytics techniques for exploring the design space of
  large-scale high-radix networks.
\newblock In {\em Proc. IEEE Cluster}, pp. 193--203, 2017.

\bibitem{ma2007situ}
K.-L. Ma, C.~Wang, H.~Yu, and A.~Tikhonova.
\newblock In-situ processing and visualization for ultrascale simulations.
\newblock In {\em Journal of Physics: Conference Series}, vol.~78, p. 012043.
  IOP Publishing, 2007.

\bibitem{mubarak2012modeling}
M.~Mubarak, C.~D. Carothers, R.~Ross, and P.~Carns.
\newblock Modeling a million-node dragonfly network using massively parallel
  discrete-event simulation.
\newblock In {\em Proc. 2012 SC Companion}, pp. 366--376. IEEE.

\bibitem{muelder2016visual}
C.~Muelder, B.~Zhu, W.~Chen, H.~Zhang, and K.-L. Ma.
\newblock Visual analysis of cloud computing performance using behavioral
  lines.
\newblock {\em IEEE TVCG}, 22(6):1694--1704, 2016.

\bibitem{pezzotti2017approximated}
N.~Pezzotti, B.~P. Lelieveldt, L.~van~der Maaten, T.~H{\"o}llt, E.~Eisemann,
  and A.~Vilanova.
\newblock Approximated and user steerable {tSNE} for progressive visual
  analytics.
\newblock {\em IEEE TVCG}, 23(7):1739--1752, 2017.

\bibitem{pierce1977causality}
D.~A. Pierce and L.~D. Haugh.
\newblock Causality\,in\,temporal\,systems: Characterization and\,a\,survey.
\newblock {\em Journal of Econometrics}, 5(3):265--293, 1977.

\bibitem{qahtan2015pca}
A.~A. Qahtan, B.~Alharbi, S.~Wang, and X.~Zhang.
\newblock A\,{PCA}-based\,change detection framework for multidimensional data
  streams: Change detection in multidimensional data streams.
\newblock In {\em Proc. ACM SIGKDD Int. Conf. Knowledge Discovery and Data
  Mining}, pp. 935--944, 2015.

\bibitem{ross2008incremental}
D.~A. Ross, J.~Lim, R.-S. Lin, and M.-H. Yang.
\newblock Incremental learning for robust visual tracking.
\newblock {\em Int. Journal of Computer Vision}, 77(1-3):125--141, 2008.

\bibitem{sanderson2018coupling}
A.~Sanderson, A.~Humphrey, J.~Schmidt, and R.~Sisneros.
\newblock Coupling the {Uintah} framework and the {VisIt} toolkit for parallel
  in situ data analysis and visualization and computational steering.
\newblock In {\em Proc. Int. Conf. High Performance Computing}, pp. 201--214.
  Springer, 2018.

\bibitem{schulz2011interpreting}
M.~Schulz, J.~A. Levine, P.-T. Bremer, T.~Gamblin, and V.~Pascucci.
\newblock Interpreting performance data across intuitive domains.
\newblock In {\em Proc. Int. Conf. Parallel Processing}, pp. 206--215. IEEE,
  2011.

\bibitem{sculley2010web}
D.~Sculley.
\newblock Web-scale k-means clustering.
\newblock In {\em Proc. Int. Conf. World Wide Web}, pp. 1177--1178. ACM, 2010.

\bibitem{turkay2017designing}
C.~Turkay, E.~Kaya, S.~Balcisoy, and H.~Hauser.
\newblock Designing progressive and interactive analytics processes for
  high-dimensional data analysis.
\newblock {\em IEEE TVCG}, 23(1):131--140, 2017.

\bibitem{turkay2018progressive}
C.~Turkay, N.~Pezzotti, C.~Binnig, H.~Strobelt, B.~Hammer, D.~A. Keim, et~al.
\newblock Progressive data science: Potential and challenges.
\newblock {\em arXiv preprint:1812.08032}, 2018.

\bibitem{maaten2008visualizing}
L.~van~der Maaten and G.~Hinton.
\newblock Visualizing data using {t-SNE}.
\newblock {\em {Journal of Machine Learning Research}}, 9:2579--2605, 2008.

\bibitem{webga2015discovery}
K.~Webga and A.~Lu.
\newblock Discovery of rating fraud with real-time streaming visual analytics.
\newblock In {\em Proc. IEEE VizSec}, pp. 1--8, 2015.

\bibitem{xu2017vidx}
P.~Xu, H.~Mei, L.~Ren, and W.~Chen.
\newblock {ViDX}: Visual diagnostics of assembly line performance in smart
  factories.
\newblock {\em IEEE TVCG}, 23(1):291--300, 2017.

\end{thebibliography}
